\documentstyle[12pt]{article}
\newcommand{\be}{\begin{eqnarray}}
\newcommand{\ee}{\end{eqnarray}}

\def\lsim{\mathrel{\rlap{\lower4pt\hbox{\hskip1pt$\sim$}}
    \raise1pt\hbox{$<$}}}               
\def\gsim{\mathrel{\rlap{\lower4pt\hbox{\hskip1pt$\sim$}}
    \raise1pt\hbox{$>$}}}               

\input{epsfig.sty}

\begin{document}

\begin{figure}[htb]

\epsfxsize=6cm \epsfig{file=logo_INFN.epsf}

\end{figure}

\vspace{-4.75cm}

\Large{\rightline{Sezione ROMA III}}
\large{
\rightline{Via della Vasca Navale 84}
\rightline{I-00146 Roma, Italy}

\vspace{0.6cm}

\rightline{INFN-RM3 98/3}
\rightline{September 1998}
}

\normalsize{}

\vspace{2cm}

\begin{center}

\LARGE{The light baryon spectrum and the exchange of pseudoscalar and vector
mesons among constituent quarks\footnote{To appear in the Proceedings of the
Third International Conference on {\em Quark Confinement and the Hadron 
Spectrum} (CONFIII), JLab (Newport News, USA), June 7-12, 1998, World 
Scientific (Singapore).}}
 
\vspace{1cm}

\large{Fabio Cardarelli and Silvano Simula}

\vspace{0.5cm}

\normalsize{Istituto Nazionale di Fisica Nucleare, Sezione Roma III,\\
Via della Vasca Navale 84, I-00146 Roma, Italy}

\end{center}

\vspace{1cm}

\begin{abstract}

\noindent The effects of the exchanges of both pseudoscalar and vector mesons
among constituent quarks on the mass spectra of light baryons are
investigated, paying particular attention to the contribution of tensor and
spin-orbit terms. It is shown that the latter ones heavily affect the
calculated spectra at variance with the empirical observation of the weakness
of the baryon spin-orbit splittings. The relativistic suppression of the
strength of the interaction among light quarks is argued to be a possible way
to reproduce the light-baryon mass spectra.      

\end{abstract}

\newpage

\pagestyle{plain}

\section{Introduction}

\indent The Constituent Quark Model is known to be a phenomenological model
able to reproduce the hadron mass spectra. Within this model the constituent
quarks ($CQ$'s) are the only relevant degrees of freedom in mesons and
baryons, all the other degrees of freedom being frozen in the $CQ$ mass
($m_i$) and interaction. The baryon wave function $\Psi$ is therefore
eigenfunction of a Schroedinger-type equation, viz. $\hat{H} \Psi = M \Psi$,
where $\hat{H} = \hat{T} + \hat{V}$ with $\hat{T} = \sum_{i=1}^3
\sqrt{|\vec{p}_i|^2 + m_i^2}$ being the kinetic term and $\hat{V} =
\hat{V}_{conf} + \hat{V}_{s.r.}$ the interaction term, given by a long-ranged
confining part $\hat{V}_{conf}$ and a short-ranged part $\hat{V}_{s.r.}$
responsible for the baryon mass hyperfine splittings.

\indent The confining potential is usually derived from a Lorentz-scalar
interaction and, as suggested by the spectroscopy and also by lattice $QCD$
calculations, it can be taken linearly dependent on the quark-quark distance
$r_{ij} \equiv |\vec{r}_i - \vec{r}_j|$, namely $\hat{V}_{conf} \rightarrow
\hat{V}_s = \sum_{i<j} b \cdot r_{ij}$, where $b$ is the string tension. As
for the short-ranged part of the quark-quark potential, two choices exist in
the literature, based on two alternative mechanisms of boson exchange among
$CQ$'s: the one-gluon-exchange ($OGE$) and the pseudoscalar Goldstone-boson
exchange ($GBE$).

\vspace{1cm}

\section{The $OGE$ + Confinement Potential}

\indent As it is well known, the non-relativistic reduction of the vector
$OGE$ interaction leads to a Coulomb-like potential
 \be
    \hat{V}_{s.r.} \rightarrow \hat{V}_v = \sum_{i<j}  
    \frac{\alpha_s}{r_{ij}} ~{\bf F}_i \cdot {\bf F}_j
 \ee
where $\alpha_s$ is the strong coupling constant and ${\bf F}_i$ is 
the color operator for quark $i$, being $\langle {\bf F}_i \cdot {\bf F}_j
\rangle = -2/3 ~ (-4/3)$ for baryons (mesons). The combination of the linear
confinement and the Coulomb-type term gives the so-called Cornell-type
potential, which was firstly  proven to be successful in reproducing the mass
spectra of heavy quarkonia. A similar success holds in the light-quark sector
too, provided a spin-spin interaction term is added in order to take into
account spin splittings (like, e.g., the $\pi - \rho$ and $N - \Delta$
splittings), namely
 \be
    \label{spin}
    \hat{V}^{(v)}_{ss} = \frac{2}{3m_i m_j} \Delta \hat{V}_v(r_{ij}) 
    \vec s_i \cdot \vec s_j
 \ee

\indent Since the spin-spin term (\ref{spin}) formally arises at first order
in the $1 / m^2$ expansion of the vector $OGE$ interaction, full consistency
requires that spin-orbit and tensor terms at the same $1 / m^2$ order have to
be considered as well, viz.
 \be
    \label{vectorLS}
    \hat{V}^{(v)}_{ls} & = & \frac{1}{r_{ij}} \frac{d \hat{V}_v}{d r_{ij}}
    \left[\frac{(\vec r_{ij} \times \vec p_i) \cdot \vec s_i}{2 m_i^2} -
    \frac{(\vec r_{ij} \times \vec p_j) \cdot \vec s_j}{2m_j^2} +
    \frac{(\vec r_{ij} \times \vec p_i) \cdot \vec s_j}{m_i m_j} -
    \frac{(\vec r_{ij} \times \vec p_j) \cdot \vec s_i}{m_i m_j} \right]
    ~~~~ \\
    \label{vectens}
    \hat{V}^{(v)}_{tens} & = & \frac{1}{m_i m_j} \left[ \frac{1}{r_{ij}}
    \frac{d \hat{V}_v}{d r_{ij}} - \frac{d^2 \hat{V}_v}{d^2 r_{ij}} \right]
    \left[\frac{(\vec s_i \cdot \vec r_{ij})(\vec s_j \cdot \vec
    r_{ij})}{r^2_{ij}} - \frac{1}{3} \vec s_i \cdot \vec s_j \right]
 \ee

\indent However, once the above correction terms are considered, a flaw
appears in the Confining + $OGE$ model: the strength of the vector spin-orbit
term is too large with respect to the one required by the light-baryon
spectroscopy. This is known as the spin-orbit puzzle, which was solved by
Isgur and coworkers {\cite {Isgur}} ~ i) by partially compensating the vector
spin-orbit term with the Thomas-Fermi precession spin-orbit term arising from
the scalar confining interaction, i.e. 
 \be
   \label{scalarLS}
   \hat{V}^{(s)}_{ls} = - \frac{1}{2 r_{ij}} \frac{d \hat{V}_s}{d r_{ij}}
   \left[\frac{(\vec r_{ij} \times \vec p_i) \cdot \vec s_i}{ m_i^2} -
   \frac{(\vec r_{ij} \times \vec p_j) \cdot \vec s_j}{m_j^2} \right]
 \ee
and ~ ii) by using the relativistic factors $\sqrt{m_i m_j / E_i E_j}$ in the
interquark potential, which yield a significant suppression of the strenght
in case of light quarks. The effective Hamiltonian model developed in ref.
{\cite {Isgur}} was very successful: it reproduces a large amount of both
meson and baryon experimental masses and solves the spin-orbit puzzle.
Nevertheless, a residual problem still remains in the generally good picture
given by the Isgur model: negative-parity states are below positive-parity
ones in clear contrast to the observation.

\vspace{1cm}

\section{The $GBE$ + Confinement Potential}

\indent The $GBE$ potential among $CQ$'s has been recently applied to the
calculation of the light-baryon spectra in ref. {\cite {Glozman}}. The
non-relativistic reduction of the $GBE$ interaction leads to a  potential of
the form $\hat{V}_{ps} = \hat{V}^{octet} + \hat{V}^{singlet}$ with
 \be
    \label{octet}
    \hat{V}^{octet} & = & \left[ \sum_{a=1}^3 V_{\pi}(r_{ij}) \lambda_i^a
    \cdot \lambda_j^a + \sum_{a=4}^7 V_{\kappa}(r_{ij}) \lambda_i^a \cdot
    \lambda_j^a + V_{\eta}(r_{ij}) \lambda_i^8 \cdot \lambda_j^8 \right] 
    \vec{\sigma}_i \cdot \vec{\sigma}_j \\
    \label{singlet}
    \hat{V}^{singlet} & = & \frac{2}{3} V_{\eta'}(r_{ij}) \vec{\sigma}_i \cdot
    \vec{\sigma}_j
 \ee
where $\lambda_i^a$ is a flavor operator for quark $i$ and $V_M$ ($M = \pi,
\kappa, \eta, \eta'$) are radial Yukawa-like functions
 \be
    \label{yukawa}
    V_M(r_{ij}) = \frac{g^2_{Mqq}}{4\pi} \frac{1}{12 m_i m_j} \vec{\sigma}_i
    \cdot \vec{\sigma}_j \left[\mu_M^2 \frac{e^{-\mu_M r_{ij}}}{r_{ij}} -
    \Lambda_M^2 \frac{e^{-\Lambda_M r_{ij}}}{r_{ij}} \right]
 \ee
The second term in the squared braket of eq. (\ref{yukawa}) is a smearing
function, depending on a cutoff parameter $\Lambda_M$, and $g_{Mqq}$ is the
quark-meson coupling costants.

\indent The $GBE$ model of ref.{\cite{Glozman}} predicts baryon masses in
quite good agreement with the experimental data. In particular, due to the
presence of the flavor-dependent factor $\vec{\lambda}_i^F \cdot
\vec{\lambda}_j^F$, the $GBE$  model is able to yield the correct ordering
among positive and negative parity states. However, as in the $OGE$ case, the
full consistency requires to consider {\em all} the terms at the same $1 /
m^2$ order appearing in the non-relativistic reduction of the interaction.
Thus, for the $GBE$ interaction both the pseudoscalar tensor term
 \be
   \label{pstens}
   \hat{V}^{(ps)}_{tens} = \frac{1}{m_i m_j} \left[  \frac{d^2
   \hat{V}_{ps}}{d r^2_{ij}} - \frac{1}{r_{ij}} \frac{d \hat{V}_{ps}}{d
   r_{ij}} \right] \left[\frac{(\vec s_i \cdot \vec r_{ij})(\vec s_j \cdot
   \vec r_{ij})}{r^2_{ij}} - \frac{1}{3} \vec s_i \cdot \vec s_j \right]
 \ee
and the scalar spin-orbit term (\ref{scalarLS}) have to be considered. These
terms were not included in the calculations of ref. {\cite{Glozman}} and we
will refer to the potential used in ref. {\cite{Glozman}} as the $Graz$
version of the $GBE$ interaction. We now briefly address the following
question: if and to what extent the general goodness of the $Graz$ model
picture is preserved when the hyperfine terms (\ref{scalarLS}) and
(\ref{pstens}) are included in the effective Hamiltonian.

\vspace{1cm}

\section{An Investigation of the Full $GBE$ Interaction}

\indent We have calculated light-baryon masses in case of both the Isgur and
$GBE$ potentials by expanding the baryon wave function in a truncated set of
harmonic oscillators and using the Raleigh-Ritz principle to determine the
corresponding variational coefficients. We have checked that the results of
refs. {\cite{Isgur}}$^-${\cite{Glozman}} are correctly reproduced. Some of
our results for the $Graz$ potential are shown in the third column of Table 1.
Then, we have introduced the pseudoscalar tensor term (\ref{pstens}) of the
$GBE$ interaction with no addition of free parameters with respect to ref.
{\cite{Glozman}}, and we have solved the corresponding Schroedinger equation.
The results obtained are shown in the fourth column of Table 1, labelled as
{\em Graz + tensor}. It can be noticed that the $N - \Delta$ splitting is
significantly lowered, while the $N - Roper$, $N - F_{15}$, $D_{13} - S_{11}$
and $S^*_{11} - S_{11}$ splittings are increased, so that the overall
agreement between predicted and experimental masses is destroyed. The last
step has been to add also the scalar spin-orbit term (\ref{scalarLS}) and
again no free parameters have been introduced. The results are shown in the
last column of Table 1, labelled as {\em full GBE}. It can be seen that even
more dramatic modifications are introduced in the predicted spectrum: the $N
- \Delta$ and $N - F_{15}$  splittings appear to be largely underestimated,
while the $S^*_{11} - S_{11}$ is strongly overestimated and, in addition, the
$D_{13} - S_{11}$ splitting becomes too big, so that the spin-orbit puzzle
reappears in the full $GBE$ model.

\begin{center}

{\bf Table 1}. $N^* - N$ mass splittings (all values are in MeV).
\begin{tabular} {||c ||c ||c ||c ||c ||}
\hline
$N - N^*$ & exp. {\cite {PDG}} & $Graz$ & $Graz + tensor$ & $full ~ GBE$ \\
\hline \hline
$N-\Delta$ & $294 \pm 2$ & $295 \pm 6$ & $217 \pm 21$ & $252 \pm 19$\\
\cline{1-5} \cline{1-5}
$N-Roper$ & $502^{+30}_{-10}$ & $528 \pm 11$ & $568\pm 20$ & $557 \pm 18$\\
\cline{1-5} \cline{1-5}
$N-D_{13}$ &  $582^{+10}_{-5}$ & $596 \pm 6$ & $651 \pm 11$ & $555 \pm 11$\\
\cline{1-5} \cline{1-5}
$N-S_{11}$ & $ 597^{+20}_{-15}$ & $596 \pm 6$ & $604 \pm 12$ & $665 \pm 12$\\
\cline{1-5} \cline{1-5}
$N-S^*_{11}$ & $712^{+30}_{-10}$ & $698 \pm 6$ & $850 \pm 16$ & $1044 \pm 19$
\\
\cline{1-5} \cline{1-5}
$N-F_{15}$ &  $792^{+10}_{-5}$ & $781 \pm 9$ & $854 \pm 13$ & $606 \pm 16$
\\
\hline
\end{tabular}

\end{center}

\vspace{1cm}

\section{Conclusions}

\indent In conclusion, we have shown that a constituent quark model based on
the complete $1 / m^2$ first-order expansion of the pseudoscalar Goldstone
boson exchange interaction leaves the spin-orbit problem unsolved in the
light-baryon spectrum. Some ideas to overcome this problem may be considered,
like: ~ (a) to add the exchange of vector mesons;  ~ (b) to introduce the
relativistic suppression factors $\sqrt{m_i m_j / E_i E_j}$ in the $GBE$
interaction, as it has been already done for the $OGE$ case by Isgur and
coworkers (both in the baryon and meson sectors).

\indent Option (a) comes from the observation that the tensor interaction due
to the vector meson exchange (eq. (\ref{vectens})) has opposite sign to the
corresponding pseudoscalar interaction (eq. (\ref{pstens})). In fact we have
checked that, by {\em excluding} all spin-orbit terms, the experimental mass
splittings can be reproduced (i.e., vector mesons compensate pseudoscalar
mesons in the tensor term). However, this is obtained at the price of having
the quark - vector meson coupling of the same order of the quark -
pseudoscalar meson one (which is not a welcome result at all). Moreover,
the vector meson exchange exhibits a spin-orbit term with opposite sign
with respect to the spin-orbit Thomas-Fermi precession term associated to the
scalar confinement (see eqs. (\ref{vectorLS}) and (\ref{scalarLS})).
Nevertheless, we have found no compensation among them, mainly because the
confining spin-orbit term is long-ranged, while vector mesons are massive.
Thus, we can state that vector meson exchange cannot adequately help the
$GBE$ model in reproducing the light-baryon spectrum.

\indent Option (b) appears much more promising, because the relativistic
suppression factors lower significantly the contribution of the spin-orbit
and tensor terms in case of light quarks. Such momentum-dependent factors
represent phenomenologically the effects of the non-locality in the effective
quark-quark potential. A throughout investigation of this point is in progress
and the results will be reported elsewhere.

\end{document}